\documentclass[12pt]{article}
\usepackage{epsf}
\usepackage{color}

\usepackage{graphicx}

\begin{document}
\begin{center}
{\bf Binding Energies and Dissociations of $J/\psi$, $\Upsilon$,
and Toponium in Quark-Gluon Plasma}
\end{center}
\vspace*{1cm}
\begin{center}
{\bf Sidi C. Benzahra, Arthur A. Evans, Robert A. Porter}
\end{center}
\begin{center}
{\it California Polytechnic State University}
\end{center}
\begin{center}
{\it San Luis Obispo, California 93407, USA}
\end{center}

\vspace*{1cm}
\begin{abstract}
\noindent In this work, we calculate the binding energies of the
$J/\psi$, the $\Upsilon$, and the Toponium in quark-gluon plasma.
We also calculate the temperature of the quark-gluon plasma at
which these quarkonia dissociate.  We approach the problem
from a quantum mechanical point of view. We use the variational method and the screening Debye potential
in order to obtain these binding energies.\\

\end{abstract}

{\bf I. Introduction}\\

Since heavy quarks at hadron colliders can be produced, the
quarkonia such as $c\bar{c}$, $b\bar{b}$, and $t\bar{t}$ can also
be produced. Experiments at LHC will eventually reach an energy
that will lead to a large number of $t\bar{t}$ events. When we
collide heavy ions to produce the quark-gluon plasma, we will
produce the quarkonia in the process, which some of it will get
embedded inside the quark-gluon plasma and dissociate due to the
hot temperature. The binding energies of the quarkonia will get
smaller when the quarkonia get inserted in the quark-gluon
plasma. For example, the binding energy of free $\Upsilon$ is
about 800 MeV, but it decreases to 10 MeV once we immerse the
$\Upsilon$ in a 150 MeV QGP [1]. The quarkonia get pumped up, so
to speak, to a lower binding energy due screening. We shouldn't
just look at the effect of screening on the binding energy, we
should also consider the effect of the absorption of gluon by
the quark, and the effect of the collision Q+g $\longrightarrow$ Q+g.
In our model we are looking at the effect of screening only, and
we agree that another mechanism would involve absorption by
gluons, or collision of a gluon
and a quark. Other mechanism could be the subject of another investigation.\\
\newpage

{\bf II Description}\\

In the high temperature deconfined phase, the quark-antiquark free
energy $V_{q\bar{q}}$, which is the Debye potential with inverse
screening length $m_{el}$, is given by [2]:
\begin{equation}
{V_{q\bar{q}}}=-{{4}\over{3}}{{\alpha_{s}}\over{r}}e^{-m_{e}r}
\end{equation}
Using the variational method the binding energy is
\begin{equation}
E={{<\psi|H|\psi>}\over{<\psi|\psi>}}
\end{equation}

Since the Debye potential is spherically symmetric, and similar to
Coulomb potential, we choose a hydrogen-like trial wave function
\begin{equation}
|\psi \rangle = Ne^{-r/a_{T}}. \label{trialw1}
\end{equation}
This trial wave function gives us a normalization term
\[{ \langle \psi|\psi \rangle }=\pi N^{2} a_{T}^{3}, \]
where $N$ is just a constant and $a_{T}$ is a variational
parameter.  Calculating the value
\[ { \langle \psi|H|\psi \rangle }= 4 \pi N^{2} \left [ {{a_{T}}\over{4
m_{Q}}}-{{4}\over{3}} \alpha_{s}{{a_{T}^{2}}\over{(2+m_{el}
a_{T})^{2}}} \right ], \]

we find the variational estimate of the binding energy of the two
quarks in the meson to be:

\begin{equation}
{E}= {1 \over {m_{Q} a_{T}^{2}}} - {16 \over 3}{\alpha_{s}} {{1}
\over {a_{T}(2 + m_{el} a_{T})^{2}}}\, . \label{binding-energy}
\end{equation}

One can see that the binding energy in Eq.(\ref{binding-energy})
is written in terms of the observables, $m_{Q}, m_{el},
\alpha_{s}$, and the parameter $a_{T}$ which we can eliminate
through the variational method.  Taking the derivative of the
energy with respect to the parameter $a_{T}$ and making it equal
to zero, we get:

\[ {{dE}\over{da_{T}}}={{32 \alpha_{s} m_{el}}\over{3 a_{T} (2+a_{T}m_{el})^{3}}}+{{16 \alpha_{s}}\over{3 a_{T}^{2}
(2+a_{T}m_{el})^{2}}}-{{2}\over{a_{T}^{3} m_{Q}}}=0 .\]

Minimizing the energy, we can find the parameter $a_{T}$ and then
plug it back into the energy. But the binding energy is related to
the inverse screening length, which is related to the temperature.
There is a relationship between the inverse screening length, the
number of flavors in the quark-gluon plasma, and the temperature
of the quark-gluon plasma. If we increase the temperature of the
plasma, the quarks and the gluons will become very active and some
of them will eventually get in between the quark and the antiquark
of the meson and cause more screening.  So the screening increases
with the temperature, and the quark and the antiquark can barely
``feel'' each other. This relationship can be algebraically noted
[2] in the following equation
\begin{equation}
{m_{el}^{2}} = {1 \over {3}} g^{2} (N + {{N_{f}} \over {2}}) T^{2}
\, , \label{kapusta1}
\end{equation}
where $N=3$ from the SU(N) group, g is the dimensional coupling
constant of the field strength, $F_{a}^{\mu \nu} =
\partial^{\mu}A_{a}^{\nu}-\partial^{\nu}A_{a}^{\mu}-g f_{abc}A_{b}^{\mu}A_{c}^{\nu}$, and $N_{f}=3$ is the number
of light flavors, which are the up, the down, and the strange quarks.\\
The uncertainties in using the leading perturbation of
Eq.(\ref{kapusta1}) relies on the second order, or the leading
correction. There has been an interest in computing the leading
correction to Eq.(\ref{kapusta1})[3,4]. It is known that this
correction cannot be computed perturbatively in non-Abelian
gauge[5]. The $O(g^{2}T)$correction to the inverse screening
length receives contributions from fundamentally non-perturbative
physics associated with the interactions, at high temperature, of
magnetic gluons with momenta of order $g^{2}T$ [6].

We will also use the temperature-dependent running coupling
constant of QCD, which is given by [3]
\begin{equation}
{{g^{2}} \over {4 \pi}} = {{12 \pi} \over {({11N -2N_{f}}) \rm ln
\left({T^{2}}/\Lambda^{2} \right)}} \, .
\end{equation}
This equation explicitly displays asymptotic freedom:
$g^2\rightarrow 0$ as $T \rightarrow \infty .$ We notice that
there is no intrinsic coupling ``constant'' on the right hand side
of this equation. The only free parameter of the theory on the
right side of this equation is the QCD energy scale, $\Lambda$,
whose numerical value is dependent on the gauge and on the
renormalization scheme chosen. If we choose the QCD energy scale
to be $\Lambda=50$ MeV, and
 $N_{f}=3$, and $N=3$ we get

\begin{equation}
{{g^{2}} \over {4 \pi}} = {{6 \pi} \over {27 \rm ln \left(T/50 \rm
MeV \right)}} \, . \label{kapusta2}
\end{equation}
Now combining Eq.(\ref{kapusta1}) and Eq.(\ref{kapusta2}) we get a
simple relationship between the inverse screening length and the
temperature.

\begin{equation}
m_{el}={{2 \pi T}\over{\sqrt{3 \rm ln(T/50)}}}\, . \label{inverse}
\end{equation}

Since the inverse-screening length, $m_{el}$, in Eq.(\ref{inverse}) is found to be 
written in terms of the temperature only, and the quarkonia binding energy of
Eq.(\ref{binding-energy}) is found to be written in terms of
$m_{el}$, we can now, to certain extent, calculate the binding
energy of any specific quarkonium we choose, knowing the
temperature of the medium. We have other observables in the
binding energy, of course, such as the mass of the quark and
the coupling constant.  But these observables are necessary and
need to be there. For example, the mass of the quark in
Eq.(\ref{binding-energy}) is tied to the quarkonia
we choose.\\
\newpage
Using $\alpha_{s}(m_{c})=0.30 \pm 0.02$ [7] for the $J/\psi$ and Plotting the binding energy in
terms of the temperature, we get:\\
\includegraphics[scale=0.5]{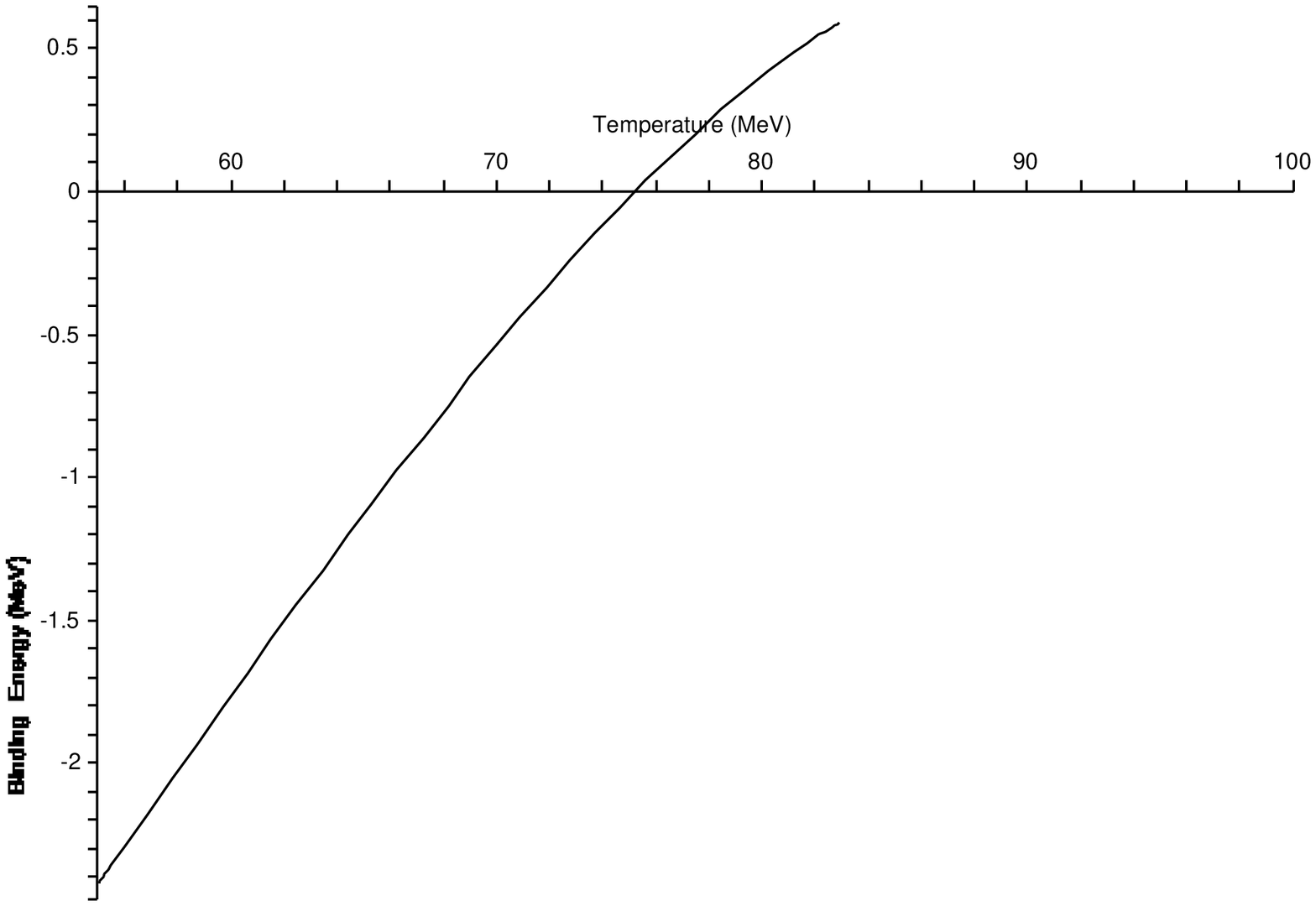}\\
When we try to look for the binding energy of the $J/\psi$ in quark-gluon plasma, we find that its curve 
doesn't cross the zero-line energy, but goes over it instead, indicating that the $J/\psi$ is unbound 
already.  But when we get too curious and we lower the $\Lambda_{QCD}$ to 30 MeV--this was done just 
to get the curve to cross the zero-line energy--we find the $J/\psi$ to be unbound at the 75 MeV temperature. This might indicate that the $J/\psi$ got dissociated before even the quark-gluon plasma was formed.
In the literature, the quark-gluon plasma materializes at a temperature of about 100 MeV or over, which makes
us believe that the $J/\psi$ has broken down before even thermalization. Our interest here is not to know 
exactly at what temperature the $J/\psi$ dissociates, but to have an idea of where the range of that 
temperature might be. We tweaked the $\Lambda_{QCD}$ just to have an idea about that temperature and not
know exactly what it is.\\
\newpage
Using $\alpha_{s}(m_{b})=0.2325 \pm 0.0044$ [8] for the $\Upsilon$ and Plotting the binding energy in
terms of the temperature, we get:\\
\includegraphics[scale=0.5]{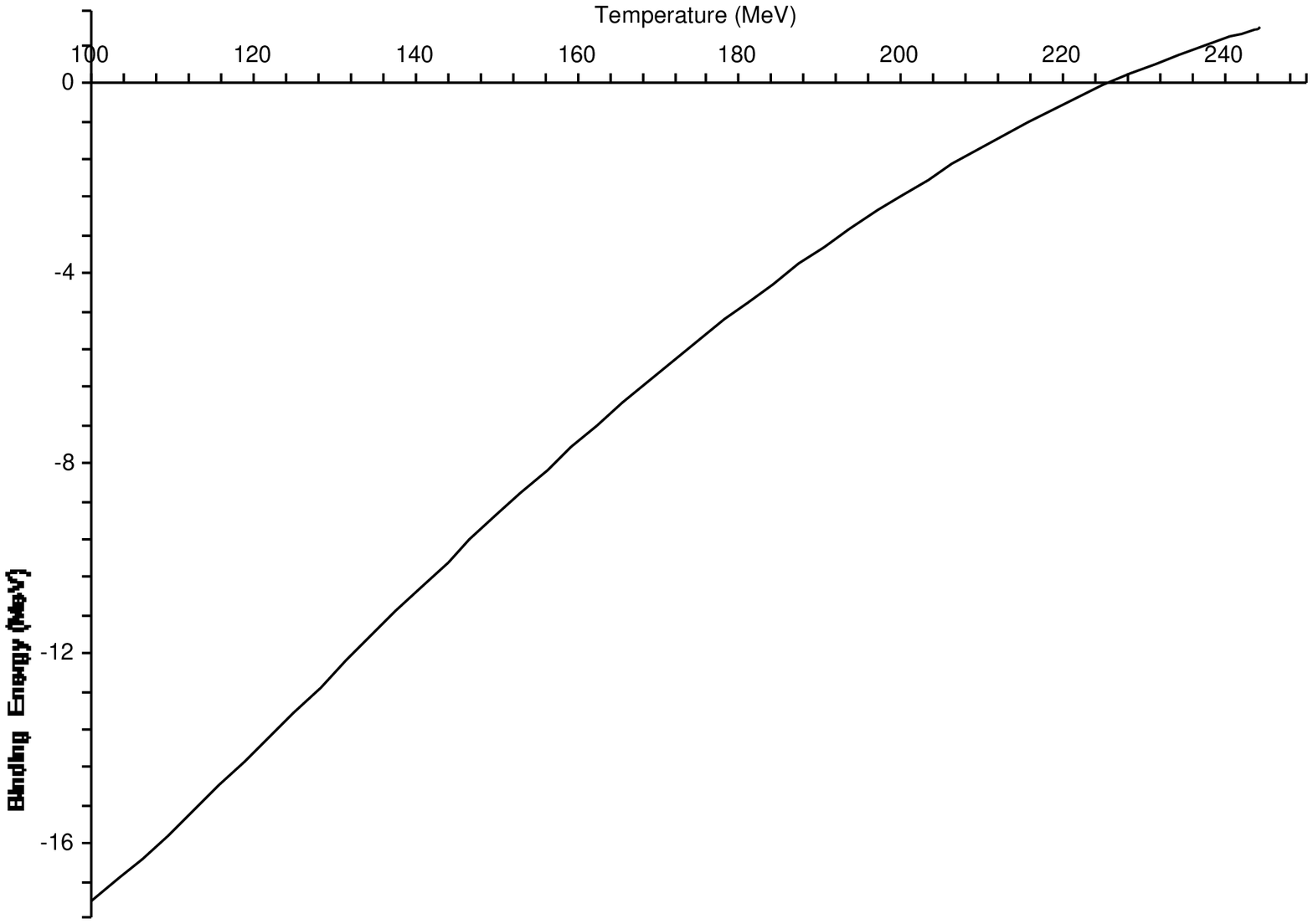}\\

In the graph just above, one can see that $\Upsilon$ dissociates at a temperature of about 225 MeV. This 
temperature is convenient and can be achieved through experiment. Since $\Upsilon$ states in a 
quark-gluon plasma are also sensitive to the color screening effect [9, 10, 11], the study of the upsilon 
meson suppression in high energy heavy ion collisions can be used as a signature for the
quark-gluon plasma as well.  Because the binding energy of $\Upsilon$ is larger than that of $J/\psi$, 
the critical energy density at which $\Upsilon$ is dissociated in the quark-gluon plasma is also 
higher [12]. One therefore expects to see the effects of the quark-gluon plasma on the production of 
$\Upsilon$ only in ultra-relativistic heavy ion collisions, such as at the RHIC and the LHC.  As in the 
case of $J/\psi$, one needs to understand the effects of $\Upsilon$ absorption in hadronic matter 
in order to use its suppression as a signal for the quark-gluon plasma in heavy ion collisions.\\
\newpage

Using $\alpha_{s}(m_{t})=0.32$[13], and Plotting the binding energy in
terms of the temperature, we get:\\
\includegraphics[scale=0.5]{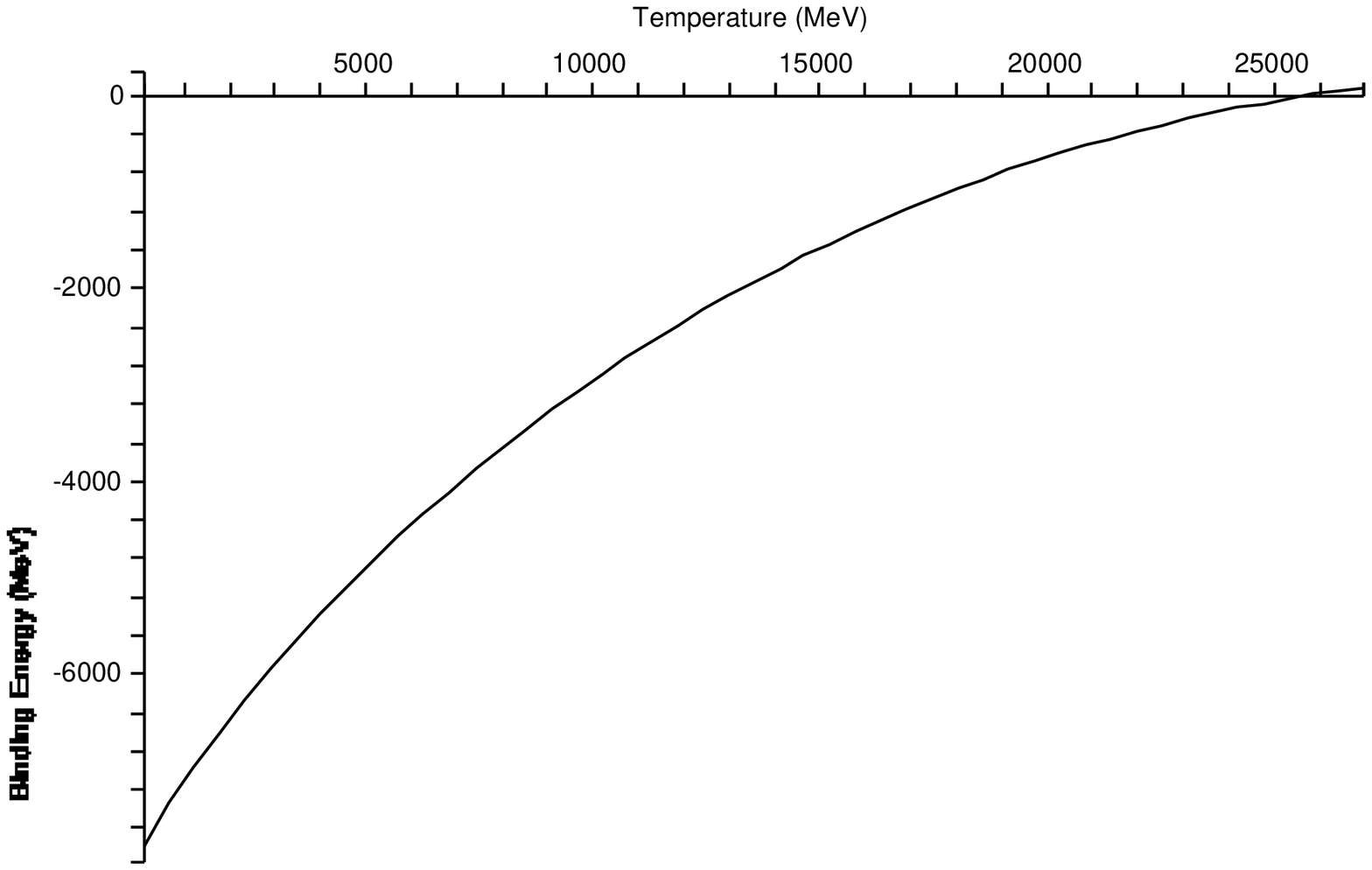}\\
We can see in the graph above that the toponium dissociates at the high temperature of 25 GeV. As of now, 
this temperature is unachievable in the lab, for it is 25 times larger than the temperature LHC is trying 
to reach in the near future.\\  

{\bf III Conclusion}\\
We state that $\Upsilon$ dissociates at 225 MeV quark-gluon plasma and the toponium dissociates at 25 GeV.
We are not satisfied with our result in regard to the $J/\psi$, because we changed $\Lambda_{QCD}$ 
considerably to get the result of 75 MeV. We know that $\Lambda_{QCD}$ is not arbitrary. Once we specify how 
$\Lambda_{QCD}$ should be determined, it is fixed.\\    

{\bf References}\\

[1] S.Benzahra and B. Bayman, hep-ph/0307271\newline [2] J.I.
Kapusta, Finite Temperature Field Theory (Cambridge University
Press, Cambridge, 1989).\newline [3] J.I. Kapusta, Phys. Rev. D
{\bf 46}, Number 10, 4749 (1992).\newline [4]K. Kajantie and J.I.
Kapusta, Phys. Lett. {\bf 110B}, 299 (1982)\newline [5]S.
Nadkarni, Phys. Rev. {\bf D22}, 3738 (1986); A.K.Rebhan, Phys.
Rev. {\bf D48}, R3967 (1993)\newline 
[6]P. Arnold and L.G. Yaffe Phys. Rev. {\bf D52}, 7208-7219 (1995)\newline
[7]M. Consoli and J.H. Field, Phys. Rev. D $\bf 49$ (1994) 1293-1301. \newline
[8] Matthias Jamin and Antonio Pich, Nucl. Phys. B $\bf 507$ (1997) 334-352.\newline
[9] T. Matsui and H. Satz, Phys. Lett. B $\bf 178$ (1986) 416. \newline 
[10] S. C. Benzahra, Phys. Rev. C $\bf 61$ 064906 (2000).\newline   
[11] For recent reviews, see, e.g., R. Vogt, Phys. Rept. $\bf 310$ (1999) 197; 
H. Satz, Rept. Prog. Phys. $\bf 63$ (2000) 1511.\newline 
[12] F. Karsch, M.T. Mehr and H. Satz, Z. Phys. C $\bf 37$ (1988) 617. \newline
[13]A. L. Macpherson and B. A. Campbell, Phys. Lett. B $\bf 306$ (1993) 379-385.\newline

\end{document}